\documentclass[twocolumn,aps,superscriptaddress,nofootinbib,floatfix,linenumbers]{revtex4}

\usepackage{amsfonts}
\usepackage{float}

\usepackage{placeins}
\usepackage{graphicx}
\usepackage[hidelinks]{hyperref}
\usepackage{color,amsmath,amssymb,bm}
\usepackage{tensor}
\usepackage[cmyk]{xcolor}
\usepackage{ulem}

\newcommand{\Occc}{$\Omega_{ccc}$ }
\newcommand{\Occ}{$\Omega_{cc}$ }
\newcommand{\Xicc}{$\Xi_{cc}^{++}$ }
\newcommand{\Xic}{$\Xi_{c}$ }
\newcommand{\pt}{$p_T$ }
\newcommand{\Dzer}{$D^0$ }
\newcommand{\Lc}{$\Lambda_c$ }
\newcommand{\Pb}{\textit{PbPb} }
\newcommand{\Kr}{\textit{KrKr} }
\newcommand{\Ar}{\textit{ArAr} }
\newcommand{\OO}{\textit{OO} }

\begin{document}
\newcommand{\cen}{\begin{math}0 \! - \! 10\%\ \end{math}}

\title [mode = title]{Multi-charmed and singled charmed hadrons from coalescence: yields and ratios in different collision systems at LHC}

\author{Vincenzo Minissale}
\email{vincenzo.minissale@dfa.unict.it}
\affiliation{Department of Physics and Astronomy "E. Majorana", University of Catania, Via S. Sofia 64, 1-95123 Catania, Italy}
\affiliation{Laboratori Nazionali del Sud, INFN-LNS, Via S. Sofia 62, I-95123 Catania, Italy}

\author{Salvatore Plumari}
\email{salvatore.plumari@dfa.unict.it}
\affiliation{Department of Physics and Astronomy "E. Majorana", University of Catania, Via S. Sofia 64, 1-95123 Catania, Italy}
\affiliation{Laboratori Nazionali del Sud, INFN-LNS, Via S. Sofia 62, I-95123 Catania, Italy}

\author{Yifeng Sun}\email{sunyfphy@sjtu.edu.cn}
\affiliation{School of Physics and Astronomy, Shanghai Key Laboratory for Particle Physics and Cosmology, and Key Laboratory for Particle Astrophysics and Cosmology (MOE), Shanghai Jiao Tong University, Shanghai 200240, China}

\author{Vincenzo Greco}
\email{greco@lns.infn.it}
\affiliation{Department of Physics and Astronomy "E. Majorana", University of Catania, Via S. Sofia 64, 1-95123 Catania, Italy}
\affiliation{Laboratori Nazionali del Sud, INFN-LNS, Via S. Sofia 62, I-95123 Catania, Italy}

\begin{abstract}
We study the production of charmed and multi-charmed hadrons in ultra-relativistic Heavy Ion Collisions coupling the transport approach for charm dynamics in the medium to an hybrid hadronization model of coalescence plus fragmentation. In this paper, we mainly discuss the particle yields for single charmed and multi-charmed baryons focusing mainly on the production of $\Xi_{cc}$ and $\Omega_{ccc}$. We provide first predictions for \Pb collision in \cen centrality class and then we explore the system size dependence through \Kr, to \Ar and \OO collisions, planned within the ALICE3 experiment. In these cases, a monotonic behavior for the yields emerges which can be tested in future experimental data.
We found about three order of magnitude increase in the production of \Occc in \Pb collisions compared with the yield in small collision systems like \OO collisions. 
Furthermore, we investigate the effects on the \Occc particle production and spectra coming from the modification of the charm quark distribution due to the different size of the collision systems comparing also to the case of thermalized charm distributions. These  results suggest that observation on the \Occc spectra and their evolution across system size can give information about the partial thermalization of the charm quark distribution as well as to its wave function width.
\end{abstract}

\maketitle
\section{Introduction}

Many probes have been proposed to investigate the properties of the matter created in ultra-Relativistic Heavy Ion collisions (uRHICs). A preminent one is the heavy quark hadron production; all the observables related to them have been considered one of the most useful probes to characterize the quark-gluon plasma. Due to their large masses Heavy quarks, namely charm and bottom, are considered as a solid probe to characterize the QGP phase \cite{Dong:2019unq,He:2022ywp,Andronic:2015wma}. They are produced by pQCD processes with a formation time $\tau_{0}<0.08 fm/c\ll\tau_{QGP}$ that permits to probe also the strong electromagnetic and vortical fields expected in the initial stage of the collision \cite{Das:2016cwd, Jiang:2022uoe, Chatterjee:2018lsx}.  On the other hand, the large mass implies a larger thermalization time w.r.t. light counterpart and appears currently to be comparable to the one of the QGP itself \cite{Dong:2019unq,Scardina:2017ipo} and HQs can probe the whole evolution of the plasma. Furthermore, they are expected to conserve memory of the history of the system evolution and the final hadrons can keep information about the out-of-equilibrium initial conditions.
Two main observables have been studied in uRHICs for HF hadrons: the heavy mesons nuclear modification factor $R_{AA}(p_T)$ \cite{STAR:2006btx, PHENIX:2005nhb, ALICE:2015vxz} , and the elliptic flow,
$v_2(p_T)$ \cite{PHENIX:2006iih, Abelev:2014dsa}. 
These observables have been studied in order to extract the heavy flavour transport coefficients and understand the HQs dynamics in QGP from a theoretical point of view \cite{vanHees:2007me,Gossiaux:2008jv,Alberico:2011zy,Uphoff:2012gb,Lang:2012nqy,Song:2015ykw,Cao:2015hia,Das:2015ana,Sun:2019fud,Cao:2018ews,Rapp:2018qla}. 
Recently, further efforts have been done to extend the analysis to higher order anisotropic flows $v_n$ \cite{Beraudo:2021ont, Katz:2019fkc, Ke:2018tsh,Plumari:2019hzp,Sambataro:2022sns} that can provide more constraints on the extraction of the transport coefficients.
The coalescence mechanism is one of the possible description for the hadronization process present in the Quark-Gluon Plasma, it is able to explain the $p_T$ baryon to meson spectra and the splitting of elliptic flow of light mesons and baryons produced in heavy ion collisions at top RHIC energies \cite{Greco:2003xt,Fries:2003vb,Greco:2003mm,Fries:2003kq,Molnar:2003ff} as well as at LHC \cite{Minissale:2015zwa}. In the last decades coalescence models were extended to include finite width to take into account for off-shell effects \cite{Ravagli:2007xx,Ravagli:2008rt,Cassing:2009vt} which however preserve the main features of the initial modeling like the enhancement of baryon production and an approximate quark number scaling of the elliptic flow $v_2$.
In particular, for open heavy flavor, the hadronization
by coalescence play an important role to determine the $R_{AA}(p_T)$ and the elliptic flow $v_2(p_T)=\langle cos (2\phi) \rangle$ of D meson affecting the evolution of $D_s$ \cite{Das:2015ana,Greco:2003vf,Scardina:2017ipo,Song:2015ykw,Rapp:2018qla,He:2022ywp,Citron:2018lsq}. 
Furthermore, coalescence approaches have predicted an unexpected large $\Lambda_c/D^0 \sim 0.5-1$ in AA collisions \cite{Oh:2009zj,Plumari:2017ntm}, that has been 
 recently observed at RHIC energies \cite{STAR:2019ank} and in
$pp, pA, AA$ collisions at LHC \cite{ALICE:2021bib,ALICE:2020wfu,ALICE:2020wla}. Such a large ratio appears to be a strong violation of the universality of the fragmentation function seen in elementary collision systems, where
the observed $\Lambda_c/D^0$ is of $O(10^{-1})$ \cite{Lisovyi:2015uqa}.
A large baryon over meson ratio in the charm sector, as observed in AA collision, is compatible with hadronization by coalescence \cite{Oh:2009zj,Plumari:2018vae,Cao:2019iqs}. Furthermore, very recently, the Catania coalescence model have correctly predicted the ratio of $\Lambda_c/D^0$, $\Xi_c/D^0$
and $\Omega_c/D^0$ even in pp collisions \cite{ALICE:2021psx,Minissale:2020bif,ALICE:2022cop}.
At the same time, in the heavy quark sector, theoretical studies have shown that the investigation on the bottom quarks can provide further insight on the HQ thermalization looking at $R_{AA}$, anisotropic flows and hadronization mechanism of $B$ mesons and $\Lambda_b$ \cite{Liu:2016ysz,Sambataro:2023tlv,Beraudo:2021ont}.
In this paper we are concentrating our focus on the production of the multi-charmed hadrons, which have two or three charm as constituent quarks. The study of these probes is, first of all, a natural extension of the investigation done for single-charmed hadrons, moreover it can provide further information about the hadronization mechanism, that should be  much more sensitive to the charm quark features in the QGP medium. The production of multicharmed hadrons are part of the physical motivation of the ALICE3 proposal for HI-LHC  \cite{ALICE:2022wwr}.
First observations of multi-charmed baryons was reported in 2002 by SELEX collaboration at Fermilab for $\Xi_{cc}^+$ \cite{SELEX:2002wqn}. Recently, LHCb collaboration has observed \Xicc in pp collisions at top LHC energies in two different decay channels \cite{LHCb:2017iph,LHCb:2022rpd}.
First theoretical studies on the multi-charmed production were made with statistical hadronization model in \cite{Becattini:2005hb}, and more recently an increasing interest has grown about the topic, with different calculations present in literature \cite{Cho:2019syk,Andronic:2021erx,He:2014tga,Zhao:2016ccp,Yao:2018zze}.
The multi-charmed hadrons, as well as the exotic states, e.g. $T_{cc}^+$, and pentaquarks, has been also indicated as  interesting for future experimental developments and investigations \cite{ALICE:2022wwr}.
The paper is organized in the following way: in Section \ref{section:Coal}, we describe our hybrid hadronization model by coalescence plus fragmentation. In Section \ref{section:Fireball}, we discuss the collision systems characteristics and parton distribution setup for our calculations. In Section \ref{section:yield_PbPb}, we present our results for the single-charmed and multi-charmed yields in \Pb collision at LHC energies compared with Statistical Hadronization Model (SHM), and the effect on the production coming from microscopical details of our model. In Section \ref{section:yield_AA}, we study the multi-charm production varying the collision systems, i.e. \Pb, \Kr, \Ar and \OO, and the role of non-equilibrium behaviour of charm quark distribution. Finally, in Section \ref{section:Conclusion}, we give our conclusions.
\section{Hybrid hadronization by coalescence and fragmentation}
\label{section:Coal}
Coalescence models have been widely applied as a mechanism of hadronization in HICs. These models were successful for the explanation of the constituent-quark number scaling of the elliptic flow $v_2(p_T)$ in $AA$ collisions for light hadrons and the large baryon-to-meson ratios for light hadron production at “intermediate” $p_T$ \cite{Fries:2008hs}.
In recent years, for the HF hadron chemistry in $AA$ collisions, the charmed hadrons production has been
investigated within these models predicting a large $\Lambda_c/D^0$ \cite{Oh:2009zj,Plumari:2017ntm,Cho:2019lxb,He:2019vgs}.
 This approach is suitable also to provide results in different collision systems, and it was proposed as an alternative approach of hadronization in very small collisions systems like  $pp$ collisions,  assuming that also in this systems a small QGP droplet can be
formed where the charm quarks, produced in perturbative processes, can hadronize via recombination with light thermal partons present in the medium\cite{Minissale:2020bif}. It was shown that this hadronization mechanism can explain in a natural way the $p_T$ dependence of different charmed hadrons ratio like $\Lambda_c/D^0$, $\Xi_c/D^0$, $\Omega_c/D^0$ at top LHC energies.

In this section, we recall the basic elements of the coalescence model developed
in \cite{Greco:2003mm,Greco:2003vf,Fries:2003kq,Fries:2003vb}
and based on the Wigner formalism. 
The momentum spectrum of hadrons formed by coalescence of quarks can be written as:
\begin{eqnarray}
\label{eq-coal}
\frac{dN_{H}}{dyd^{2}P_{T}}=g_{H} \int \prod^{N_{q}}_{i=1} \frac{d^{3}p_{i}}{(2\pi)^{3}E_{i}} p_{i} \cdot d\sigma_{i}  \; f_{q_i}(x_{i}, p_{i}) \\ 
\times C_{H}(x_{1}...x_{N_{q}}, p_{1}...p_{N_{q}})\, \delta^{(2)} \left(P_{T}-\sum^{n}_{i=1} p_{T,i} \right) \nonumber
\end{eqnarray}
with $g_{H}$ we indicate the statistical factor to form a colorless hadron from quarks and antiquarks with spin 1/2. For mesons with spin-0 the statistical factors $g=1/36$ gives the probability that two random quarks have the right colour, spin, isospin to match the quantum number of the considered mesons.
For baryons with spin-1/2 the statistical factors is $g=1/108$.
The $d\sigma_{i}$ denotes an element of a space-like hypersurface, 
while $f_{q_i}$ are the quark (anti-quark) phase-space distribution functions for i-th quark (anti-quark). $C_H(...)=\mathcal{N} f_H(...)$ with $\mathcal{N}$ a normalization factor; while 
$f_{H}(x_{1}...x_{N_{q}}, p_{1}...p_{N_{q}})$ is the Wigner function which  
describes the spatial and momentum distribution of quarks in a hadron.


Following the Refs. \cite{Oh:2009zj,Plumari:2017ntm,Greco:2003vf}
we adopt for the Wigner distribution function a Gaussian shape in space and momentum, 
\begin{equation}
 f_H(...)=\prod^{N_{q}-1}_{i=1} 8\exp{\Big(-\frac{x_{ri}^2}{\sigma_{ri}^2} - p_{ri}^2 \sigma_{ri}^2\Big)}
\end{equation}
where $N_{q}$ is the number of constituent quarks.\\ 
Notice that $\mathcal{N}$ has been fixed to guarantee
that in the limit $p \to 0$ all the charm hadronize by coalescence in a heavy hadron.
This is imposed by requiring that the total coalescence probability for charm quarks gives $\lim_{p \to 0} P^{tot}_{coal}=1$.
It has been shown, by other studies, that the inclusion of missing charm-baryon states \cite{He:2019vgs} 
or the variation of the width of the D meson wave function \cite{Cao:2019iqs,Cho:2019lxb}, can permit 
that all the zero momentum charm quarks can be converted to charmed hadrons.
The relative coordinate are evaluated going into the CM  frame of the particles involved in the process and are defined as follows.
For mesons the relative coordinates $(r_1, p_{r1})$ are given by,
\begin{eqnarray}
r_1=|\vec{x}_1-\vec{x}_2|, ~ p_{r1}={|m_2\vec{p}_1-m_1\vec{p}_2|\over m_1+m_2},
\end{eqnarray}  
while for baryons are defined as
\begin{eqnarray}
    r_{1}=\frac{|\vec{x}_1-\vec{x}_2|}{\sqrt{2}}, ~ p_{r1}=\sqrt{2} \; {|m_2\vec{p}_1-m_1\vec{p}_2|\over m_1+m_2},
\end{eqnarray}  
and $r_{2}$, $p_{r2}$  are given by
\begin{eqnarray}
r_{2}&=& \sqrt{2 \over 3} \;  \left| {  {m_1\vec{x}_1+m_2\vec{x}_2\over m_1+m_2}}-\vec{x}_3 \right| , \nonumber\\
p_{r2}&=& \sqrt{3 \over 2} \;  {|m_3(\vec{p}_1+\vec{p}_2)-(m_1+m_2)\vec{p}_3|\over m_1+m_2+m_3} ,
\end{eqnarray}  

The $\sigma_{ri}$ are the covariant widths, that can be related to the oscillator
frequency $\omega$ by $\sigma_{ri}=1/\sqrt{\mu_i \omega}$ where $\mu_i$ are the reduced masses
\begin{eqnarray} 
\mu_1=\frac{m_1 m_2}{m_1+m_2}, & \, & \mu_2= \frac{(m_1+ m_2)m_3}{m_1+m_2+m_3}.
\end{eqnarray}
In our calculations the masses of light and heavy quarks have been fixed to $m_{u,d}\!=\!300$ MeV, $m_{s}\!=\!380$ MeV, $m_{c}\!=\!1.5$ GeV.
The Wigner function 
for the heavy meson has only one parameter which is the width $\sigma_{r 1}$, while for baryons there are two parameters $\sigma_{r 1}$ and $\sigma_{r 2}$  that are related by the oscillatory frequency $\omega$ through the reduced
 masses by
\begin{equation} \label{Eq:sigma_omega}
\sigma_{p i}=\sigma_{ri}^{-1}=1/\sqrt{\mu_{i} \omega}
\end{equation}
The widths of the Wigner function $f_H$ are fixed by using the relation with the size of the hadron and in particular to the root-mean-square charge-radius of the hadron, $\langle r^2\rangle_{ch}= \sum_{i=1}^{N} Q_i\langle(x_i-X_{cm})^2\rangle$ with $N=2,3$ for mesons and baryons respectively; see ref.\cite{Plumari:2017ntm,Minissale:2020bif} for single-charmed mesons and baryons.
The mean square charge radius of mesons and single charmed baryons used as reference come from quark model \cite{Hwang:2001th,Albertus:2003sx}. 
The widths for heavy hadron used in this work are shown in Table \ref{table:param}.
The corresponding mean square charge radii evaluated from these widths have values within the uncertainties coming from the quark models calculation mentioned above.
In table \ref{tab:widths_multi} we report the widths and radii for multi-charm hadrons.  To fix the widths values of the Wigner function in the case of multi-charmed hadrons (\Xicc and $\Omega_{ccc}$) we start from the frequency $\omega$ of the single charmed hadrons, \Xic and $\Omega_c$ respectively; and then we calculate the widths with a scaling from the frequency and the new reduced masses according to Eq.\ref{Eq:sigma_omega}.
\begin{table} [ht] \label{tab:single-c widths}
\begin{center}
  \begin{tabular}{l |c c c }
    \hline
    \hline
    Meson &$\langle r^2\rangle_{ch}$ & $\sigma_{p1}$ & $\sigma_{p2}$ \\ 
    $D^{+}=[c \bar{d}]$     & 0.184   & 0.226 & --- \\
    $D_{s}^{+}=[\bar{s}c]$   & 0.083   & 0.24  & --- \\ 
    \hline
    Baryon &$\langle r^2\rangle_{ch}$ & $\sigma_{p1}$ & $\sigma_{p2}$ \\ 
    $\Lambda_c^+ =[udc]$	   & 0.15   & 0.305  & 0.502 \\ 
    $\Xi_c^+ =[usc]$	   & 0.2   & 0.291  & 0.487 \\ 
    $\Omega_c^0 =[ssc]$	   & -0.12   & 0.404  & 0.636 \\ 
\hline
\hline
\end{tabular}
\end{center}
\caption{Mean square charge radius $\langle r^2\rangle_{ch}$ in $fm^2$ and the widths parameters $\sigma_{pi}$ in $GeV$. The mean square charge radius are taken quark model \cite{Hwang:2001th,Albertus:2003sx}.}
\label{table:param}
\end{table}

\begin{table}
\begin{tabular}{c|c|c|c|c} 
\hline 
\hline
 \tiny
 & $\Xi_{c}$ & $\Omega_{c}$ & $\Xi_{cc}^{(scal. \omega)}$ &  $\Omega_{ccc}^{(scal. \omega)}$\\
\hline 
\hline
 $\sigma_{p_1}(GeV)$ & 0.262 & 0.345 & 0.317 &0.668 \\
 $\sigma_{p_2}(GeV)$ &   0.438 &  0.557 & 0.573 &  0.771  \\
 $\sigma_{r_1}(fm)$ &  0.751 &   0.572 & 0.622 & 0.295 \\
 $\sigma_{r_2}(fm)$ &   0.450 & 0.354 & 0.344 & 0.256  \\
 $\langle r^2\rangle_{ch}(fm^2)$&  0.2   &  -0.12 & 0.363 &  0.09 \\
 $\langle r^2\rangle(fm^2)$ &  0.745  &0.428  & 0.545 &  0.13 \\
 $\omega$ & $1.03e\!-\!2$ &$1.5e\!-\!2$ &$1.03e\!-\!2$ & $1.5e\!-\!2$  \\
\hline 
\end{tabular} 
\caption{Multi-charmed widths parameters $\sigma_{pi}$ in $GeV$, mean radii $\langle r^2\rangle$ in $fm^2$ and frequencies compared with single-charmed baryons.}
\label{tab:widths_multi}
\end{table}
As known from previous works on coalescence \cite{Minissale:2015zwa,Plumari:2017ntm,Cao:2015hia,Gossiaux:2009mk,Oh:2009zj} the coalescence probability decreases at increasing $p_T$, this behaviour let the  the standard independent fragmentation to be the dominant hadronization process for the production at high $p_T$. Hence, the inclusion of the hadronization by fragmentation is necessary to describe correctly the transition to the high momentum regime but does not affect significantly the yield. 
In our approach the smooth transition from low to high $p_T$ regime is given by introducing a fragmentation probability $P_{frag}(p_T)$. 
As done in Ref.\cite{Plumari:2017ntm,Minissale:2020bif}  we start from the probability that one charm quark can hadronize by coalescence and we assume that charm quarks that
do not hadronize via coalescence 
are converted to hadrons by fragmentation.
The fragmentation probability is given by $P_{frag}(p_T)=1- P^{tot}_{coal}(p_T)$, where $P^{tot}_{coal}$ is the total coalescence probability.
Notice that with the same approach used in \cite{Plumari:2017ntm,Minissale:2020bif} the coalescence probability for a charm quark to hadronize via coalescence is forced to be 1 at $p_T \approx 0$. 
The hadron momentum spectra from the charm parton fragmentation is given by:
\begin{equation}
\frac{dN_{had}}{d^{2}p_T\,dy}=\sum \int dz \frac{dN_{fragm}}{d^{2}p_T\, dy} \frac{D_{had/c}(z,Q^{2})}{z^{2}} 
\label{Eq:frag}
\end{equation}
$D_{had/c}(z,Q^{2})$ is the fragmentation function and $z=p_{had}/p_{c}$ is the momentum fraction
of heavy quarks transferred to the final heavy hadron while 
$Q^2=(p_{had}/2z)^2$ is the momentum scale for the fragmentation process.
In our calculations for charm quarks we have used the Peterson fragmentation function \cite{Peterson:1982ak}
$D_{had}(z) \propto 1/[ z [1-z^{-1}-\epsilon_c({1-z})^{-1}]^2 ]$
where $\epsilon_c$ is a free parameter that is determined assuring 
that the shape of the fragmentation function agrees with the experimental data on $p_T$ distributions for \Dzer and \Lc at $p_T>10 \rm \; GeV$
The $\epsilon_c$ parameters used here are the same as in \cite{Plumari:2017ntm}. 
We notice that at high $p_T$ the fragmentation becomes to be the dominant charm hadronization mechanism and a quark will hadronize according to the different fragmentation fractions into specific final charmed hadron channels.
The fragmentation fraction is evaluated according to PYTHIA8 ratios at high $p_T> 10 \rm \; GeV $ that are similar to the $e^++e^-$ \cite{Lisovyi:2015uqa} apart from an increase of the fraction for \Lc and moderate decrease of the fraction going to \Dzer, as already done in \cite{Plumari:2017ntm,Minissale:2020bif}.
In our calculation the multi-dimensional integrals in the coalescence formula are evaluated by using a  
Monte-Carlo method, see \cite{Greco:2003xt,Plumari:2017ntm} for more details.

\section{Fireball and parton distribution}\label{section:Fireball}
%
%
\begin{table} \label{Table:Fireball}
\centering
\begin{tabular}{c|c|c|c|c} 
\hline
\hline 
                    		& OO	& ArAr & KrKr & PbPb \\ 
\hline 
\hline
$R_{0} (fm)$ 	             &  2.76 & 3.75 & 4.9    & 6.5  \\ 
$R_{max} (fm)$ 	            & 5.2   & 7.65  & 10.1   & 14.1  \\ 
$\tau (fm)$         		&	4 & 5   & 6.2  & 8	 \\ 
$\beta_{max}$      			& 0.55	& 0.6  & 0.64   & 0.7   \\ 
$V_{|y|<0.5} (fm^3)$		& 345	& 920  & 2000   & 5000  \\ 

\hline 
\end{tabular} 
\caption{Fireball radii, lifetime, flow and volume considered in the different collision systems studied.}

\end{table}

In our calculation the bulk of particles that we assume is a thermalized system
of gluons and $u,d,s$ quarks and anti-quarks.
The partons are distributed uniformly in the transverse plane and in the rapidity range $|y_{z}|<0.5$.
The longitudinal momentum distribution is assumed to be boost-invariant in the range $y\in(-0.5,+0.5)$, and is included a radial flow with the following radial profile $\beta_T(r_T)=\beta_{max}\frac{r_T}{R}$,
where $R$ is the transverse radius of the fireball.
Partons at low transverse momentum,  $p_T<2 \,\mbox{GeV} $, are considered thermally distributed
\begin{equation}
\label{quark-distr}
\frac{dN_{q,\bar{q}}}{d^{2}r_{T}\:d^{2}p_{T}} = \frac{g_{q,\bar{q}} \tau m_{T}}{(2\pi)^{3}} \exp \left(-\frac{\gamma_{T}(m_{T}-p_{T}\cdot \beta_{T})}{T} \right) 
\end{equation}
where $m_T=\sqrt{p_T^2+m_{q,\bar{q}}^2}$ is the transverse mass.
The factors $g_{q}=g_{\bar{q}}=6$ are the spin-color degeneracy.
The presence of gluons in the quark-gluon plasma is taken into account 
by converting them to quarks and anti-quark pairs according to the flavour compositions, as assumed in \cite{Biro:1994mp,Greco:2003mm}.

\begin{figure}
\centering
\includegraphics[scale=0.31,clip]{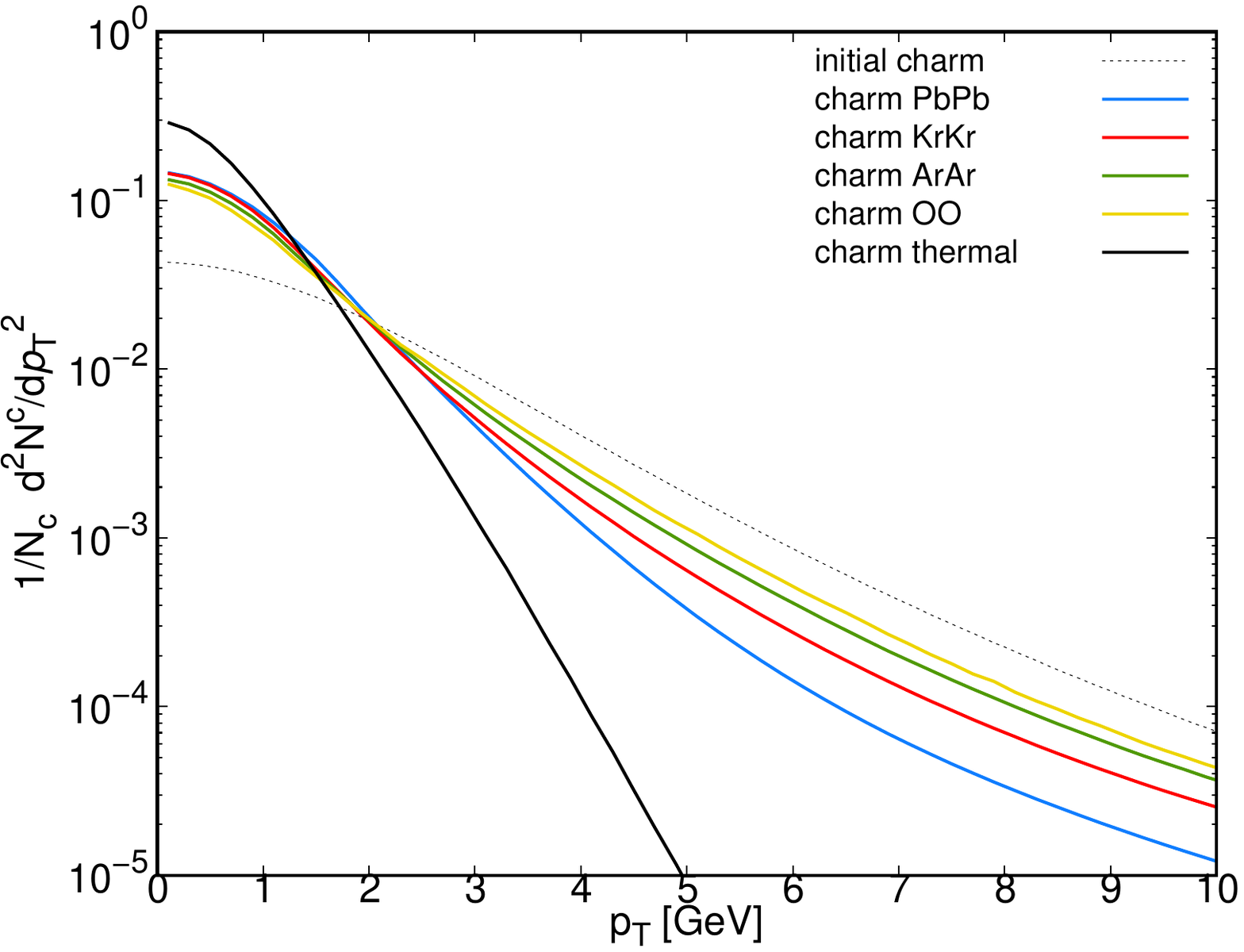}
\caption{Normalized charm distributions: from FONLL (black dashed line) before evolution, after evolution in $PbPb$ (blue solid line), $KrKr$ (red solid line), $ArAr$ (green solid line), $OO$ (gold solid line) collisions and the thermal production (black solid line)}
\label{Fig:charm_parton}
\end{figure}

The volume of the fireball in one unit of rapidity is given by $V = \pi R^2_\perp \tau$
where $R_\perp$ is the radius of the fireball taking into account the radial expansion.
We fix the radial flow and the volume by imposing the total multiplicity $dN/dy$ and the total transverse energy $dE_T /dy$ to be equal to the experimental data.
This constraints lead to values for radial flow, radius and $\tau$ as shown in Table \ref{Table:Fireball},
in quite good agreement
also with simulations in kinetic transport
approaches, and with Statistical Hadronization Model estimate for the active hadronization Volume \cite{Andronic:2021erx}. 
Such values correspond in one unity
of rapidity to fireball volume  for $PbPb$, $KrKr$, $ArAr$ and $OO$ that are, respectively, $5000 fm^3$, $2000 fm^3$, $920 fm^3$ and $345 fm^3$.
For the initial $p_T$ distribution of partons at high transverse momentum, $p_T > 2.5 GeV$,  we have considered the mini-jets that have undergone the jet quenching mechanism. As done in Ref.\cite{Plumari:2017ntm}, we have considered the initial $p_T$ distribution from pQCD calculation and the thickness function of the
Glauber model to go from pp collisions to AA ones.
The charm pair production is described by hard process
and it is described by perturbative QCD (pQCD) at NNLO.
Therefore, the starting point to compute the initial heavy quarks spectra 
at LHC collision energy of
$\sqrt{s}=5.02 \, TeV$ is by pQCD calculation. In our calculation the initial charm quark spectrum have been taken in accordance 
to the Fixed Order + Next-to-Leading Log 
(FONLL), as given in Refs. \cite{Cacciari:2005rk,Cacciari:2012ny}.
The number of charm quarks has been chosen in accordance with a charm quark cross section of $d\sigma_{c\bar{c}}/dy \sim 0.500$  that scales from $PbPb$ to $ArAr$, $KrKr$ and \OO with the mean nuclear thickness function $T_{AA}$. \cite{ALICE:2013hur}. This cross section and $T_{AA}$ scaling gives, for these calculations, a number of charmed quarks that are $N_c^{PbPb}\!\sim\!15$, $N_c^{KrKr}\!\sim\!4.35$, $N_c^{ArAr}\!\sim\!1.5$ and $N_c^{OO}\!\sim\!0.4$.\\
Finally, the charm quark distribution evolution is obtained starting from the perturbative distribution (FONLL) and solving the relativistic Boltzmann transport equations for charm quarks
scattering in a bulk medium of quarks and gluons. The scattering cross section giving a drag and diffusion corresponding to a space transport coefficient $D_s (p\to 0)$ in agreement to lQCD \cite{Dong:2019unq,Sambataro:2023tlv,Scardina:2017ipo}. In \Pb such an approach is able to provide a good prediction for $R_{AA}(p_t)$,$v_2(p_t)$ and $v_3(p_t)$ \cite{Scardina:2017ipo,Das:2015ana,Plumari:2019hzp,Sambataro:2022sns,Sambataro:2023tlv}.\\
In Fig.\ref{Fig:charm_parton} are shown the final charm quark transverse momentum distribution obtained after the evolution in QGP, in different collision systems. The shown spectra are normalized to the same number of charm quarks in order to compare them as a function of \pt. In particular  we compare these spectra with the initial FONLL distribution, the fireball lifetime decreases with the system size, therefore in smaller systems charm quarks have a thermalization time that is larger w.r.t. the fireball lifetime, 
suggesting that the high \pt distribution remains close to the non-equilibrium pQCD initial distribution. In fact, in Fig.\ref{Fig:charm_parton}, moving from larger systems, like \Pb (blue solid line), to small systems, i.e. $OO$ (gold solid line),  we observe a flattening of the 
 charm quark spectra.\\
 Furthermore, we have explored also the extreme scenario where the charm quark are fully thermalized as in Eq.\ref{quark-distr} (black solid line in Fig.\ref{Fig:charm_parton}); from one hand this case allow us to have a more direct comparison with SHM model, and from the other hand it give us the opportunity to explore the sensitivity to the charm distribution function on the multi-charmed hadrons production.

\section{Production in Pb-Pb collisions}\label{section:yield_PbPb}

\begin{figure}
\centering
\includegraphics[scale=0.29,clip]{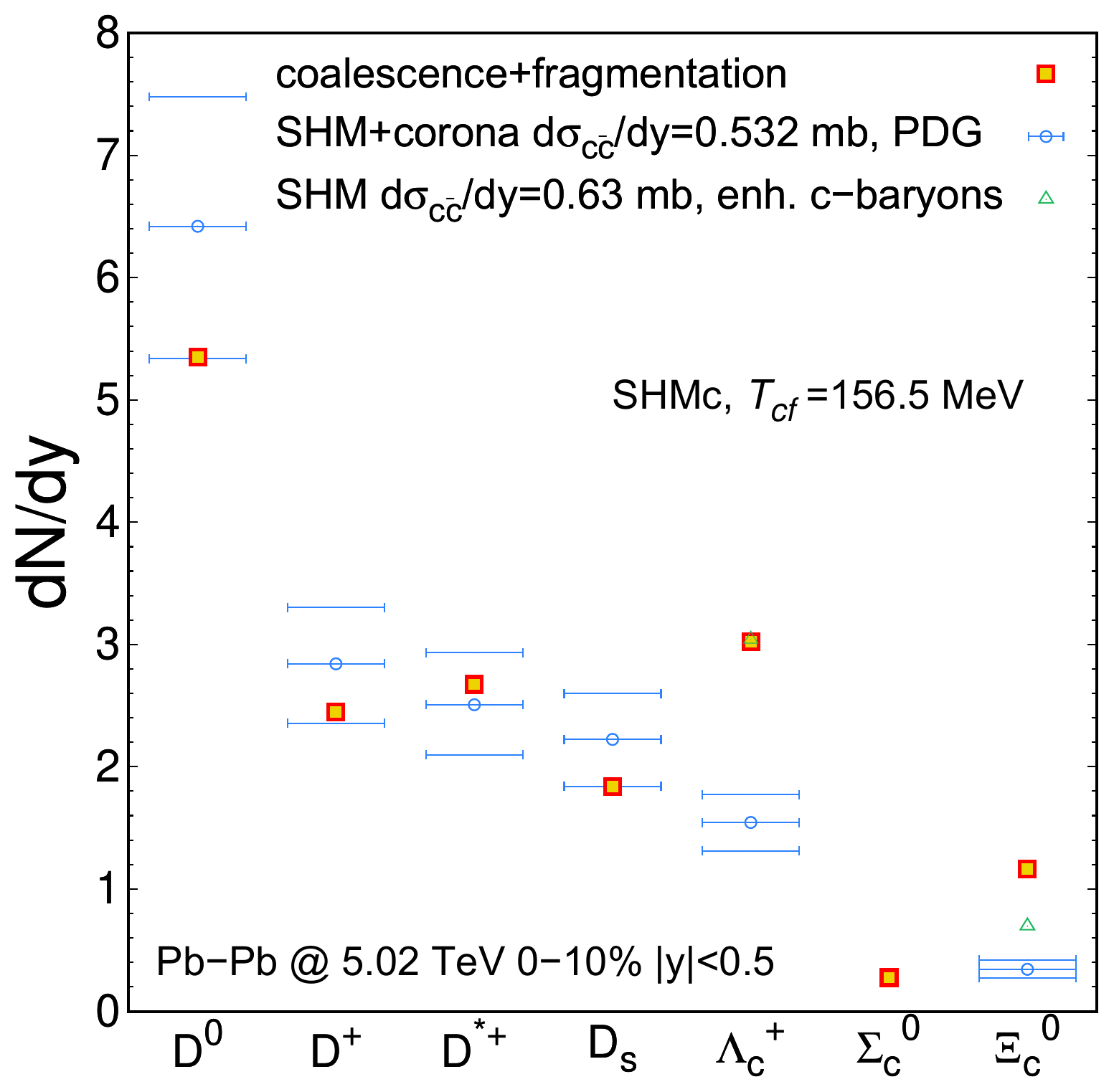}
\caption{Single-charmed hadrons yields in PbPb collision at midrapidity in \cen centrality class at $5.02 \rm \;TeV$. Our colescence plus fragmentation (red-yellow squares) results in  Comparison with SHM (blue open circles) and SHM with enhanced set of baryons (green open triangles) \cite{Andronic:2021erx}.}
\label{Fig:Yield_PbPb_singlecharm}
\end{figure}
\begin{figure}
\includegraphics[scale=0.31,clip]{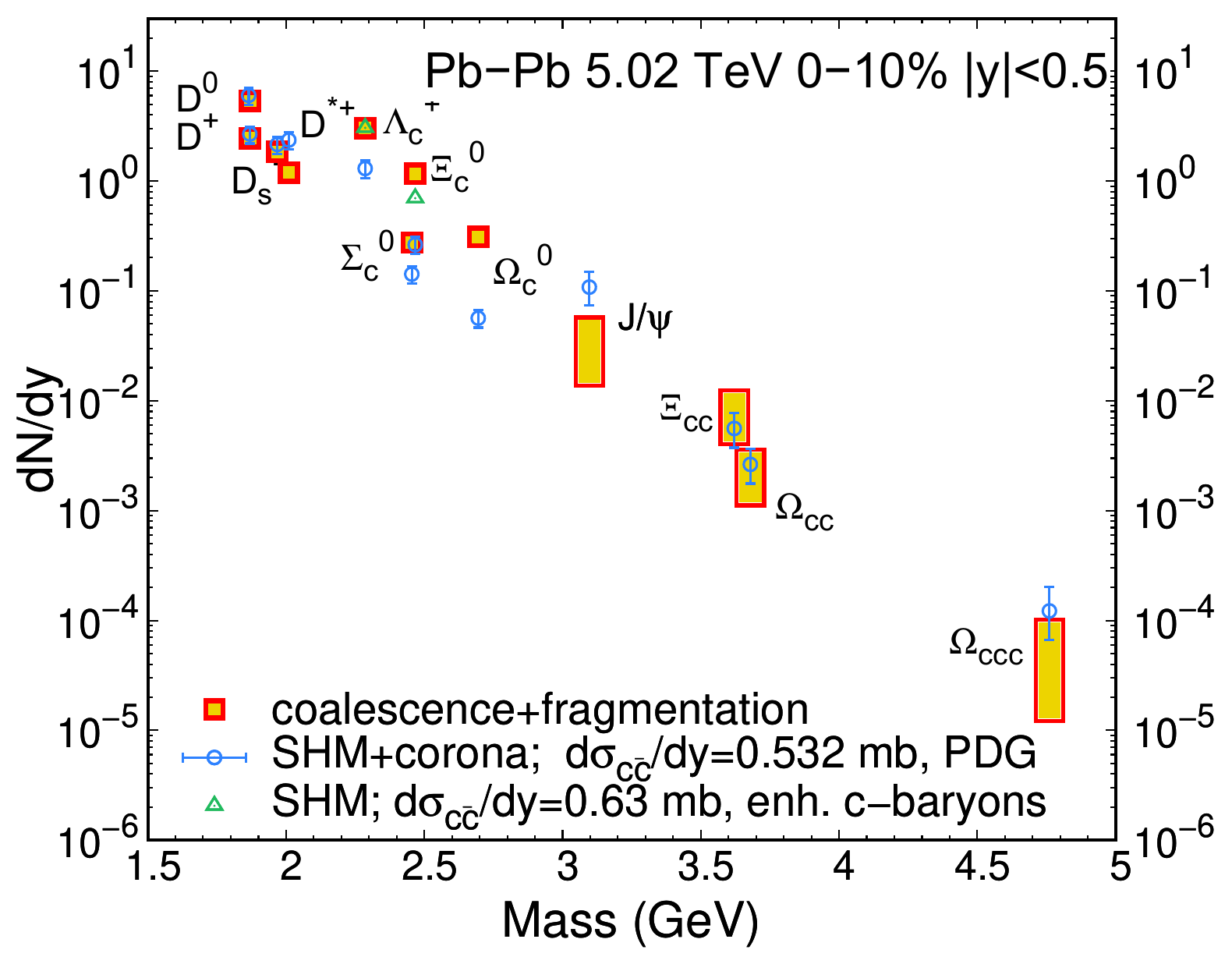}
\caption{Charmed hadrons yields in PbPb collision at midrapidity at \cen centrality class at $5.02 \rm TeV$. Our coalescence plus fragmentation (red-yellow squares) results in  Comparison with SHM (blue open circles) and SHM with enhanced set of baryons (green open triangles) \cite{Andronic:2021erx}.}
\label{Fig:Yield_PbPb_allcharm}
\end{figure}

In this section, we discuss the results for the total yields of charmed and multi-charmed hadrons using the model described in the previous section in the case of \Pb collisions at $\sqrt{s}\!=\!5.02 \rm \; TeV $. The presence of the resonance has a relevant impact because it supply a substantial contribution in addition to the ground state production. In this work we include multiple states for the different species, the complete set of considered  states is listed in Table \ref{tab:charm} for single charmed hadrons and in Table \ref{tab:multi-charm} for multi-charmed hadrons. The single-charmed resonances considered are the ones present and confirmed, at the moment, by the Particle Data Group \cite{ParticleDataGroup:2022pth}.
Following the same approach used in \cite{Plumari:2017ntm} we consider a statistical factor given by 
$[m_{H^*}/m_H]^{3/2} \times \exp {\left(-\Delta m /T\right)}$ with $\Delta m=m_{H^*}-m_H$, where
$m_{H^*}$ is the mass of the resonance.
This statistical factor is given by the Boltzmann probability to populate an excited state of mass $m+\Delta m$, at a temperature $T$.
\begin{table} [ht]
\begin{center}
\begin{tabular}{lcclr}
\hline
Meson & Mass(MeV)  & I (J) & Decay modes   & B.R. \\
\hline 
$D^+ =\bar{d}c$		& 1869 & $\frac{1}{2} \,(0)$	&\\
$D^0 =\bar{u}c$ 	& 1865 & $\frac{1}{2} \,(0)$	&\\
$D_{s}^{+} =\bar{s}c$	& 2011 & $0 \,(0)$		& \\
\hline
Resonances    \\
\hline
$D^{*+} $	& 2010 & $\frac{1}{2} \, (1)$	& $D^0 \pi^+$; $D^+ X$    & $68\%$,$32\%$ \\
$D^{*0} $	& 2007 & $\frac{1}{2} \, (1)$	& $D^0 \pi^0$; $D^0 \gamma$    & $62\%$,$38\%$\\
$D_{s}^{*+} $	& 2112 & $0 \, (1)$		& $D_{s}^+ X$ & $100\%$ \\
\hline
\hline
Baryon &  &  & &\\
\hline 
$\Lambda_c^+ =udc$	& 2286 & $0 \, (\frac{1}{2})$	&\\
$\Xi_c^+ =usc$	& 2467 & $\frac{1}{2} \, (\frac{1}{2})$	&\\
$\Xi_c^0 =dsc$	& 2470 & $\frac{1}{2} \, (\frac{1}{2})$	&\\
$\Omega_c^0 =ssc$	& 2695 & $0 \, (\frac{1}{2})$	&\\
\hline
Resonances &    \\
\hline
$\Lambda_c^+$	& 2595 & $0 \, (\frac{1}{2})$	&$\Lambda_c^+ \pi^+ \pi^-$ & $100\%$ \\
$\Lambda_c^+$	& 2625 & $0 \, (\frac{3}{2})$	&$\Lambda_c^+ \pi^+ \pi^-$ & $100\%$\\
$\Sigma_c^+$	& 2455 & $1 \, (\frac{1}{2})$	&$\Lambda_c^+ \pi$ & $100\%$\\
$\Sigma_c^+$	& 2520 & $1 \, (\frac{3}{2})$	&$\Lambda_c^+ \pi$ & $100\%$\\
$\Xi_c^{'+,0}$	& 2578 & $\frac{1}{2} \, (\frac{1}{2})$	&$\Xi_c^{+,0} \gamma$ & $100\%$\\
$\Xi_c^{+}$	& 2645 & $\frac{1}{2} \, (\frac{3}{2})$	&$\Xi_c^{+} \pi^-$, & $100\%$\\
$\Xi_c^{+}$	& 2790 & $\frac{1}{2} \, (\frac{1}{2})$	&$\Xi_c^{'} \pi$, & $100\%$\\
$\Xi_c^{+}$	& 2815 & $\frac{1}{2} \, (\frac{3}{2})$	&$\Xi_c^{'} \pi$, & $100\%$\\
$\Omega_c^{0}$	& 2770 & $0 \, (\frac{3}{2})$	&$\Omega_c^0 \gamma$, & $100\%$\\
\hline
\end{tabular}
\end{center}
\caption{Ground states of charmed mesons and baryons as well as their first excited states including their decay modes with their corresponding branching ratios as given in Particle
  Data Group \cite{Zyla:2020zbs,Agashe:2014kda}.
\label{tab:charm}}
 \end{table}
%
\begin{table} [ht]
\begin{center}
\begin{tabular}{lcc}
\hline
\hline
Baryon &  & \\
\hline 
$\Xi_{cc}^{+,++} =dcc,ucc$  	& 3621 & $\frac{1}{2} \, (\frac{1}{2})$\\
$\Omega_{scc}^+ =scc$	& 3679 & $0 \, (\frac{1}{2})$\\
$\Omega_{ccc}^{++} =ccc$	& 4761 & $0 \, (\frac{3}{2})$	\\
\hline
Resonances &    \\
\hline
$\Xi_{cc}^*$    	& 3648 & $\frac{1}{2} \, (\frac{3}{2})$ \\
$\Omega_{scc}^{*}$	& 3765 & $0 \, (\frac{3}{2})$        	\\
\hline
\end{tabular}
\end{center}
\caption{Ground states and first excited states for multi-charmed baryons. \label{tab:multi-charm}}
\end{table}
In Fig.\ref{Fig:Yield_PbPb_singlecharm} are shown the yields for single charmed hadrons in $PbPb$ collisions for \cen centrality at mid-rapidity obtained with our model of coalescence plus fragmentation (red-yellow square points). Some parameters (Temperature and Volume) in our model have been set in order to compare the results with the particle production predicted by the Statistical Hadronization Model (blue open circle points)\cite{Andronic:2021erx}. In Fig.\ref{Fig:Yield_PbPb_singlecharm} are also shown the results for SHM considering an enhanced set of charmed baryons (green open triangle points) with respect to the ones listed by the PDG, as suggested in studies on charm hadron production with statistical models \cite{He:2019vgs,He:2019tik}.
This result seems in accord with former indication for the production of charmed baryons in $pp$ and $PbPb$ collisions; where our coalescence plus fragmentation model predicts a baryon enhancement with respect to SHM. Our results are in very good agreement with recent measurements in $PbPb$ collisions at $5.02 \rm \;TeV$  with
a \Lc  production $dN^{\Lambda_c}/dy=3.28\pm 0.42(stat)\pm0.44(syst)\pm0.16(BR)$ \cite{ALICE:2021bib}.
In pp collisions the results of our coalescence plus fragmentation model 
are similar to the SHM ones, within the uncertainties, in the case of $D^0$,$D^+$,$D^*$,$D_s$, but exhibit a significant difference for \Lc and $\Xi_c$ considering statistical models with the same number of resonances  \cite{Minissale:2020bif, He:2019tik}; this behaviour seems to be confirmed also in \Pb collisions. 
In Fig.\ref{Fig:Yield_PbPb_allcharm} are shown the results for all charmed hadrons considered, adding with respect to Fig.\ref{Fig:Yield_PbPb_singlecharm} the $J/\psi$, the $\Omega_c$ and the multi-charmed baryons \Xicc, \Occ and \Occc. 
The $\Omega_c$ production in our model is about one order of magnitude smaller with respect to the \Lc yield, a results that is in line with what has been obtained in our previous work in pp collision \cite{Minissale:2020bif}, that turns out to be  larger than the statistical yield of about a factor ten.
 The multi-charmed baryons that we consider, i.e. \Xicc,\Occ and \Occc, have yields of about 3 and 5 order of magnitude smaller than the total charm quarks available in the system formed during the collision, i.e. the $dN/dy$ for \Xicc is about $(8\!\pm2)\cdot\!10^{-3}$, for \Occ is about $(0.5\!\pm0.02)\cdot\!10^{-3}$ and for \Occc  is between $(0.12\!-\!1.01)\! \cdot\!10^{-4}$, see also Table \ref{Table:Yields}.
 For multi-charmed baryons we have found a quite large sensitivity to the underlying charm \pt distribution function. If we assume a fully thermalized charm distribution we obtain an enhancement of the yields w.r.t. the realistic distribution shown in Fig.\ref{Fig:charm_parton}, this can be expected because when a thermal distribution is considered there is a large presence of charm quarks concentrated in a small momentum region at very low \pt w.r.t. the realistic distributions. This feature, facilitates the recombination mechanism, because of the larger probability to find charm quarks close in phase space in the region where is present the peak of the charm distribution function. For the multi-charmed hadrons, this property results in an enhancement in the final total yields, that is more sensitive with respect to the single-charmed hadrons because of their quark content. In fact the range shown for this particles in the plot, corresponds to the yields obtained with realistic distribution for the lower limit and the one obtained with thermal distribution for the upper limit.
In Fig.\ref{Fig:Yield_PbPb_singlecharm} and Fig.\ref{Fig:Yield_PbPb_allcharm} we compare our results for all the charmed hadrons with the production obtained with SHM (blue open circles) \cite{Andronic:2021erx}. From the comparison we observe that our model gives an enhancement for the single-charmed baryons, i.e. \Lc, \Xic and $\Omega_c$ of about a factor $\sim \!2\!-\!3$. However the baryons production becomes similar when in the SHM an enhanced set of baryon resonances is considered, as shown in figures by the green open triangles. For the multi-charmed hadrons production shown in Fig.\ref{Fig:Yield_PbPb_allcharm}, especially for \Occc SHM predicts a larger production w.r.t. the coalescence results with realistic distribution (lower limit of the band), but a similar production is obtained when the thermal distribution is used also in the coalescence model. 
In Fig.\ref{Fig:Yield_PbPb_allcharm} we have reported also the $J/\psi$ production in both dynamical and thermal charm cases, with a resulting yields that is in the range $(1\!-\!6) \cdot 10^{-2}$, about a factor 2 smaller than the SHM result and experimental measurements \cite{ALICE:2019nrq,ALICE:2023gco}. However, with our modeling we are considering that the production for the $J/\psi$ happens at the freeze-out hypersurface with temperature $T_c$, while in dynamical dissociation and recombination 
approaches consider the survival probability of $J/\Psi$ above $T_c$ \cite{Zhao:2007hh,Liu:2009nb}
that we have not included in our modeling. 
Considering a recombination model based on a dynamical approach goes quite beyond the scope of this paper that is focused in providing first predictions and system size dependence of charmed baryons.

\subsection{Sensitivity to hadron size}

\begin{figure} [b!]
\centering
\includegraphics[scale=0.4,clip]{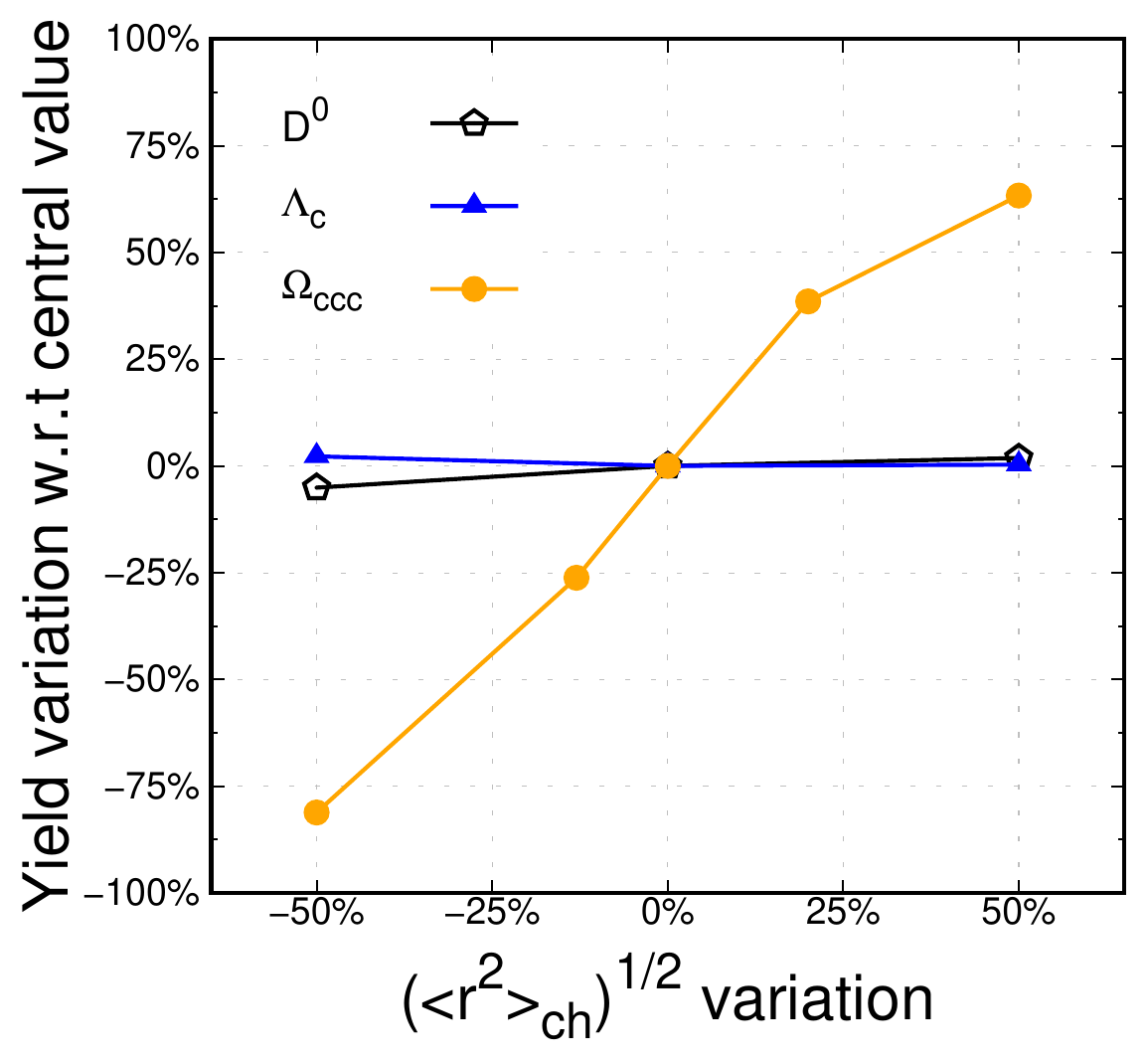}
\caption{ \Dzer (black open diamonds with line), \Lc (blue triangle with line) and \Occc (orange circles with line) yield variation at variance of $\langle r_{ch}^2\rangle^{1/2}$ variation normalized to the yield obtained with Wigner function widths present in Table \ref{table:param} and Table \ref{tab:widths_multi}}. 
\label{Fig:Occc_meanradius}
\end{figure}
In this section, we explore how the microscopic details of our hadronization model have a role on the \Occc production. In fact while for D, $D_s$, \Lc and \Xic we employ the results from the quarks model, for \Occc the mean square radius and in general the wave function is quite unknown. As discussed in the previous section the Wigner function depends only on the widths, that are directly related to the mean charge radius.
In order to obtain a quantitative estimate of the effect produced by the variation of the Wigner function widths we have performed a calculation in \Pb  varying the widths in such a way to have a resulting variation in the square root of the charged squared radius in a range between -50\% and 50\% of the established value discussed in Section \ref{section:Coal}. It is also interesting to compare the effects on the \Occc production with those for \Dzer and \Lc.
In Fig.\ref{Fig:Occc_meanradius} are shown the variation of the particle yield for \Dzer (black open diamonds with line), \Lc (blue triangle with line) and \Occc (orange circles with line) with respect to the yield obtained with the previous widths, used in this way as a baseline reference. This variation is shown as a function of the variation of $\langle r_{ch}^2\rangle^{1/2}$.
We notice that the \Dzer (and so $D^+$, $D^*$,$D_s$) and \Lc yields are almost unchanged in the two extreme cases, this behaviour can be explained recalling that this two species provide the majority of the charm hadron production; an enlargement or a shrinking of the spatial widths would normally lead to a larger or a smaller production respectively, but the imposed  charm quark conservation in conjunction with the constrained coalescence probability at zero momentum engender a compensation of the size change effect. 
The just mentioned concurrence of constraints has not a big effect on the \Occc production; in particular because, in this case, the particle production is five orders of magnitude smaller than the two aforementioned particles. As a consequence, the production of this multi-charmed particle have an impact that is negligible on the condition about the charm quark conservation. In this way, the outcome of changing the hadron radius, in \Occc case, is a larger variation of the total production w.r.t. \Dzer and \Lc; it can be quantitative described by an increase of about 60\% when the radius is increased of the 50\%, and a reduction of about 80\% when the radius is decreased  by 50\%.
Furthermore we have found for the \Xicc a very similar behaviour like the one shown for the \Occc, we don't show it in Fig.\ref{Fig:Occc_meanradius}.\\
Finally assuming that the \Occc production via fragmentation is marginal, due to the  very large mass of the baryons, this result suggests that the observations about this multi-charm particle production is very sensitive to the microscopic characteristics of the hadronization process and in particular to the wave function. This is particularly interesting because for charm quark it could be employed a potential model to compute the \Occc wave function using the heavy quark free energy from the lQCD. This would be similar to the $J/\Psi$, but \Occc production, having 3 charms, should be even more sensitive to the wave function. A seminal work in this direction can be found in \cite{He:2014tga}. Moreover the production of this multi-charmed hadrons can be  a clearer probe with respect to other charmed hadrons production, because it is partially disentangled from the effects of charm number conservation constraint. 

\section{Production evolution with collision system:Pb-Pb, Kr-Kr, Ar-Ar, O-O}\label{section:yield_AA}
\begin{figure}
\centering
\includegraphics[scale=0.32,clip]{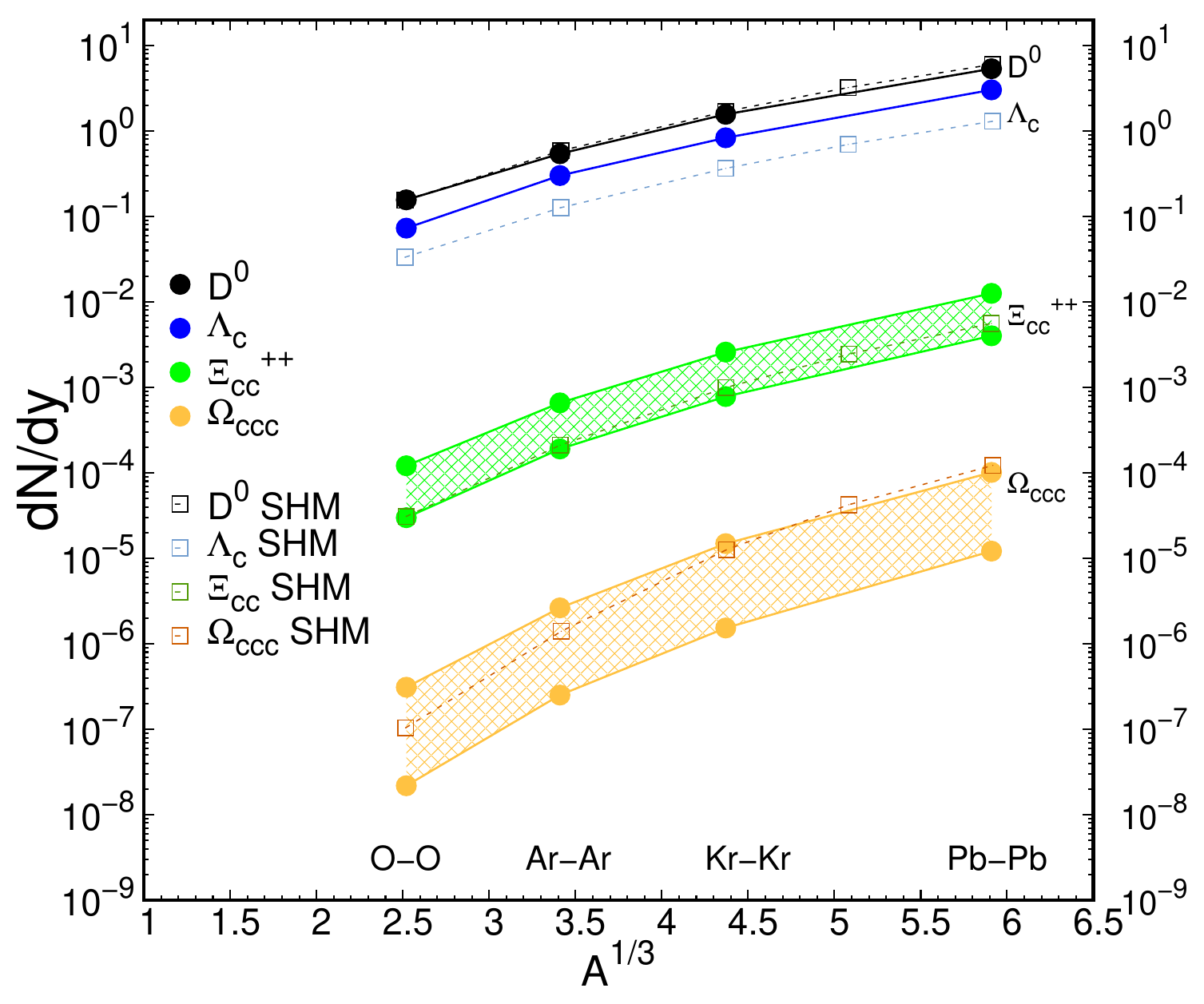}
\caption{\Dzer (black circles with line), \Lc (blue circles with line), \Xicc (green circles with line) and \Occc (orange circles with line) yields from coalescence in central PbPb, KrKr, ArAr and OO collision at midrapidity, compared to SHM model (open squares) \cite{Andronic:2021erx} }
\label{Fig:Yield_AA}
\end{figure}

\begin{figure}
    \centering
    \includegraphics[scale=0.31,clip]{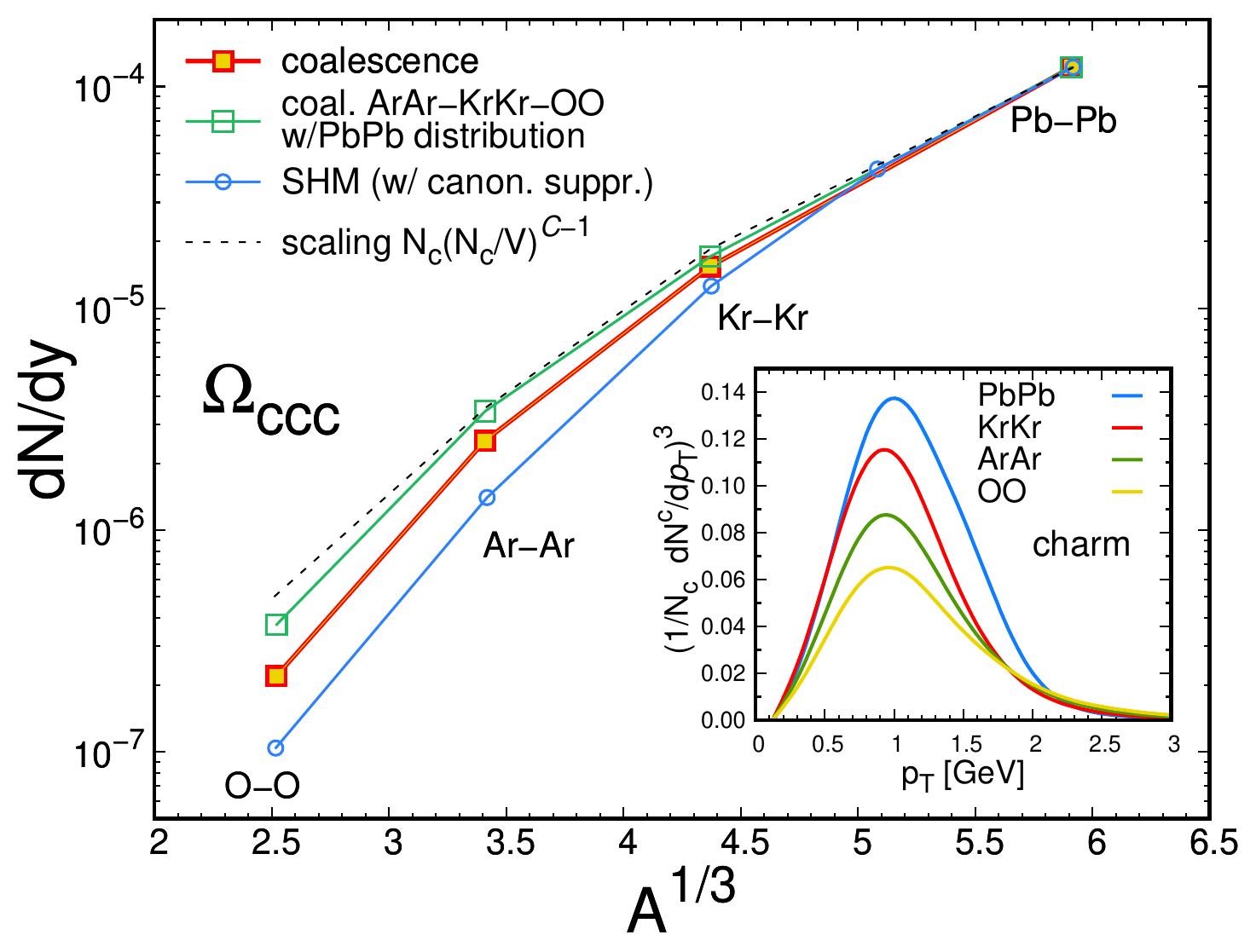}
    \caption{\Occc production in different collision systems considering: coalescence with realistic charm distributions (red-yellow squares), coalescence with fixed charm quark distributions (green open squares), the expected scaling with $N_c$ and $V$ (blasck dashed line) compared with SHM model results (blue open circle)\cite{Andronic:2021erx}. [inset] Cube of charm quark distribution in different collision systems PbPb (blue solid line), KrKr (red solid line), ArAr (green solid line) and OO (gold solid line)}
    \label{fig:Occ_syst}
\end{figure}

In this section, we want to examine the effect on the production caused by the specifics of parton distributions in different collision systems.
We have employed the model set for \Pb collision system in some other collision system, in particular in \Kr, \Ar and \OO. As discussed in Sec. \ref{section:Fireball} the fireball parameters are summarised in in Table \ref{Table:Fireball} and the charm $dN_c/dy$ scales with the $T_{AA}$ thickness function. \\
As in the case of \Pb collisions, we start from FONLL \pt distribution that evolves in a QGP medium described by a relativistic Boltzmann approach. The differences in the final \pt distribution are shown in Fig.\ref{Fig:charm_parton}. As expected, in smaller systems, the final transverse momentum charm quark distributions are flatter.
The $dN/dy$ obtained by  coalescence plus fragmentation for each species, as a function of $A^{1/3}$, are shown in Fig.\ref{Fig:Yield_AA} in comparison with SHM results and summarized in Table \ref{Table:Yields}. 
As shown, our productions scales accordingly to SHM for the single charmed particle, albeit an overall larger production of \Lc (blue circles with line) as already shown in Fig.\ref{Fig:Yield_PbPb_singlecharm}. \\
For \Occc (orange circles with line) and \Xicc (green circles with line), we evaluate the yields scaling the Wigner function widths assuming the same oscillator frequency of \Xic and $\Omega_c$  as discussed in Section \ref{section:Coal} (see Eq.\ref{Eq:sigma_omega}). We observe that our model predictions for the particle yields, have a similar but still different scaling with the decrease of the system size in comparison with SHM.\\
Notice that for the absolute yields we show a band where the upper limit comes from the assumption of fully thermalized charm quarks while the lower limit comes from the realistic simulation of partial thermalization evaluated through the Boltzmann transport approach.
In the realistic case, the \Xicc absolute production  shown in Figure \ref{Fig:Yield_AA} is close to SHM values (full square) for all the systems. Instead for the \Occc absolute production we observe a reduction for the realistic case in all systems compared to SHM results; for the fully thermalized case there is an agreement in \Pb, as discussed before and showed in Fig.\ref{Fig:Yield_PbPb_allcharm}, and an enhancement w.r.t. SHM in smaller systems.
Moreover, going back to the \Occc yield, its absolute value depends also on the assumption for the wave function width (as discussed in Sec.\ref{section:yield_PbPb}, see Fig.\ref{Fig:Occc_meanradius}). An increase of 50\% of the \Occc  charged radius $\langle r^2\rangle^{1/2}_{ch}$ can give an increment of about a 60\% for the lower limit of the band in Fig.\ref{Fig:Yield_AA}. As already mentioned a study of \Occc from lQCD potential can asses this aspect in a quite solid way. 
However, in Fig.\ref{Fig:Yield_AA} it is clear that the degree of charm thermalization play a major role. We discuss here the full thermal case for the sake of a more direct comparison to SHM. Upcoming data on \pt distribution of \Dzer, $D^+$ should put more stringent constraints on the \pt distribution, but we see from Fig.\ref{Fig:Yield_AA} that \Occc will be quite sensitive to the degree of charm thermalization. 
Furthermore, as we will discuss later in this Section, the \Occc momentum spectra will be a strong meter of the charm thermalization.\\
In general the charmed hadrons production depends on the charm quark number ($N_c$) and the system size at the freeze-out, i.e. fireball volume ($V$). 
In order to understand how the production of  multi-charmed hadrons changes with the system size, we can relate $N_c$ and $V$ to the mass number $A$. 
In first approximation, neglecting the difference in the radial flow effect that are not so different from one system to another, the volume is proportional to the mass number $V\propto A$.
The charm quark number comes from the perturbative hard processes in the initial stage of the collision, so the scaling of this quantity is expected to be proportional to $N_{coll}$. From Glauber model, in a central collision of two identical nuclei $AA$, the mean number of collisions scales as $N_{coll} \propto A^{4/3}$, then charm quark number should scales as  $N_c\propto A^{4/3}$.
From naive considerations, the coalescence mechanism has a production that is proportional to the product of the volume and the densities of the  constituent quarks involved in the hadron formation. Therefore for a charmed hadron, this scaling can be expressed as  $V\left(\frac{N_c}{V}\right)^C=N_c\left(\frac{N_c}{V}\right)^{C-1}$ where $C$ is the number of charm quarks contained by the hadron. 
Finally, considering the above relation between these quantities and the mass number, the hadron production scaling roughly results in  $N_H\propto A^{\frac{C+3}{3}}$. Such a scaling is the one expected also in SHM if canonical suppression is discarded.
We now concentrate our focus on the yield dependence as a function of $A$ in the  \Occc case, which is particularly sensitive to this scaling because of the presence of three charm quarks.
In order to disentangle the effects coming from the distribution function and the ones coming from the system size change, we have performed calculation where we consider fireball size and parameters as in Table \ref{Table:Fireball}, but employing a charm momentum distribution with the same \pt dependence of \Pb even in \Kr, \Ar and \OO.\\
In Fig.\ref{fig:Occ_syst} the \Occc  productions are shown, and for clarity's sake we have scaled all the curves in such a way to obtain that different cases has the same yields when \Pb collisions are considered.
In the above figure different cases are shown: the yields obtained with our model in the case of realistic distribution (red-yellow square with line), the expected scaling with $A$ (black dashed line), the SHM production (blue circles with line), and finally what we obtain considering for \Kr, \Ar and \OO the charm distribution used in \Pb in a fireball consistent with the reduced dimension, as said before (green open squares with line).
The last case has been realized aiming to see an hypothetical effect of the increasingly non-equilibrium of the charm quark distribution with the decreasing of the system size. 
In the inset of  Fig.\ref{fig:Occ_syst} we show the cube of the normalized charm distribution in all the systems. One can easily realize that the integration in \pt of this quantity provides a rough estimate of the effect to the \Occc yield due to the difference in momentum dependence. For example going from \Pb to \Ar there is a reduction of the production by a factor $\sim 1.7$.
Hence, in a coalescence approach, one should see a similar reduction of the \Occc yield due only to the change of the \pt charm distribution function between \Pb and other collision systems.
\\
Therefore the results shown in Fig.\ref{fig:Occ_syst} give the indication that the simple scaling $\propto A^{\frac{C+3}{3}}$ lacked information about the \pt distribution impact on the final total production. On the other hand, this scaling  is compatible with the production obtained when the charm distribution is maintained fixed, has can be seen comparing black dotted line and green solid line.
It is relevant to underline that the SHM model uses a factor that consider the canonical suppression, this factor is close to the unity for single charmed hadrons in \Pb and has a decreasing behaviour going to systems with smaller $A$, and it becomes larger increasing the charm quark content \textit{C} of the hadron  \cite{Andronic:2021erx}.
In our model this suppression factor is not present; the impact of taking into account this factor in the multi-charmed hadrons yield turn out to be an  underestimation when the system size decreases.
\begin{figure}
    \centering
\includegraphics[scale=0.38,clip]{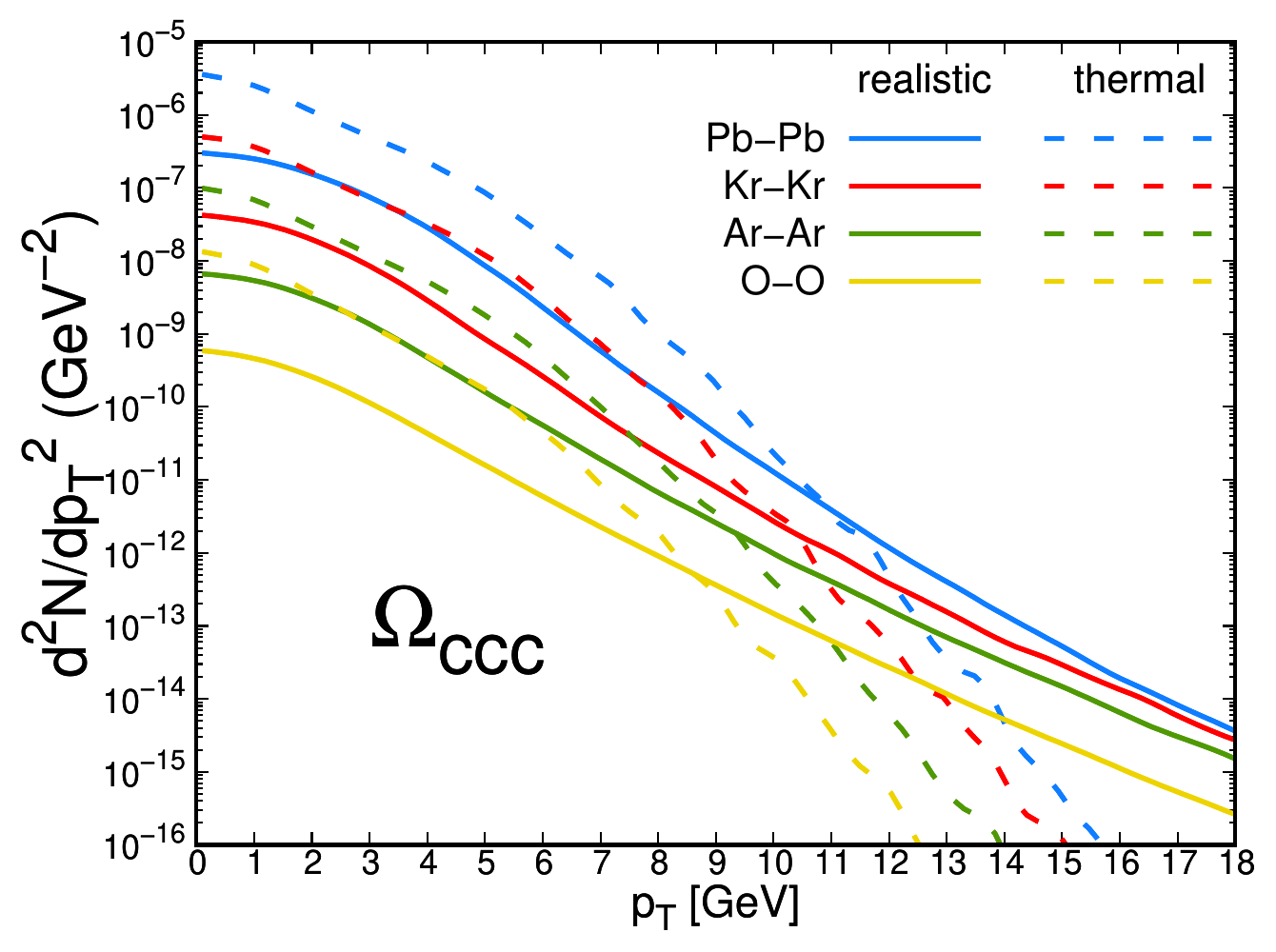}
    \caption{$\Omega_{ccc}$ spectra showed in different collision system, $PbPb$ (blue lines), $KrKr$ (red lines), $ArAr$ (green lines) and $OO$ (yellow lines). Two cases are shown, results with realistic distributions (solid lines) and thermal distributions (dashed lines).}
    \label{fig:Occc_spectra}
\end{figure}
In Fig.\ref{fig:Occc_spectra} the \Occc momentum distribution are shown in different collision systems in both realistic (solid lines) and thermal (dashed lines) cases. In Sec.\ref{section:yield_PbPb} was pointed out that the shape of thermal distribution falls down quickly w.r.t. the realistic distributions, as one should expect such a behaviour is reflected also in the \Occc spectra. Moreover, as shown for the \Occc yield in \Pb collisions in Fig.\ref{Fig:Yield_PbPb_allcharm}, the production results enhanced in the thermal case.
This behaviour should be expected looking at the  parton distribution in Fig.\ref{Fig:charm_parton}, considering that the hadron \pt is  the sum of quark momenta and the fact that thermal distribution is higher than the dynamical one below to $\sim 2 \;\rm GeV$. Hence at \pt above 6 GeV the slope of the \Occc spectra are very different and at high momentum the tail of realistic distributions result in a \pt spectrum various order of magnitude larger w.r.t. the thermal case.
\begin{figure}
    \centering
\includegraphics[scale=0.33,clip]{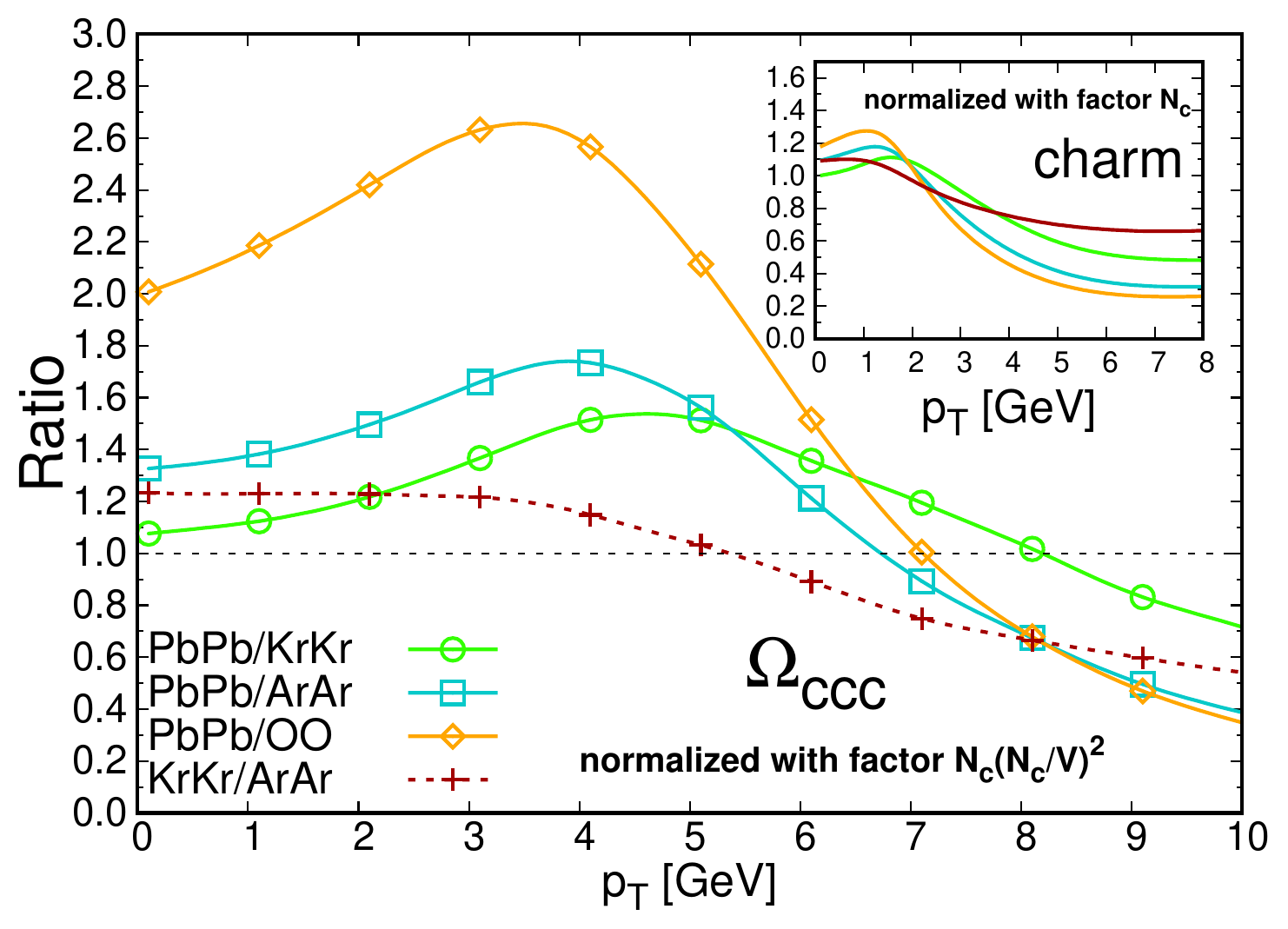}
    \caption{Ratios of $\Omega_{ccc}$ production normalized with a factor $N_c\left(N_c/V\right)^{2}$, showed in different collision system, $PbPb$, $KrKr$, $ArAr$ and $OO$. [inset] Ratio of normalized charm quark distribution in different collision systems. }
    \label{fig:Occc_ratio}
\end{figure}

Finally, in Fig.\ref{fig:Occc_ratio} the ratios between the $\Omega_{ccc}$ transverse momentum spectra are shown in different collision systems. Each spectrum is normalized with a factor $N_c\!\left(N_c/V\right)^{2}$ in order to have ratios that are comparable along the different system size. The low-\pt part, up to $4 \mbox{ GeV}$, of the \Pb ratios shows an increasing trend that reflects the presence of a stronger flow with respect to \Kr, \Ar and \OO; on the other hand the decreasing ratio at high \pt is determined by the large non-thermalization in $KrKr$, \Ar and \OO distributions. 
Notice that the \Pb/\OO ratio (orange line) with the normalization $N_c\!\left(N_c/V\right)^{2}$ has a value that is approximately 2 at low $p_T$. This effect is related to the non-thermalization in \OO, as pointed out in the discussion about Fig.\ref{fig:Occ_syst}, where the yields obtained in smaller systems with realistic distributions manifest a reduction with respect to the $A^{(C+3)/3}$ scaling.
In the inset of Fig.\ref{fig:Occc_ratio} we show the corresponding ratios of charm distribution normalized to $N_c$ in various systems, with the same colours as for the \Occc ratios. We can observe that the \Occc ratios reflect the charm ratios; the role of the hadronization is to stretch in momentum the ratio obtained at the charm quark level, so that the behaviour at charm very low \pt ($\sim\!1\rm GeV$) is transposed to the \Occc intermediate momentum region, i.e. 4-5 GeV. The ratios at charm quark level reflect the non-equilibrium effect in the evolution through the QGP as discussed for Fig.\ref{Fig:charm_parton}.
This behaviour, given the particularity of the \Occc, is much evident and sensitive with respect to similar ratios evaluated for \Dzer and \Lc. In these cases besides  the effect due to the presence of light quarks there is also a strong contribution that comes from the fragmentation, that can hide the direct effect coming from the charm distribution. Hence the $\Omega_{ccc}$ distribution might unveil direct information about the charm quark distribution with a larger sensitivity w.r.t. \Dzer or \Lc.
Summarizing, the consequence of the spectrum flattening  going from larger to smaller systems is observed as a reduction of the ratio at higher transverse momentum; producing, on the other hand, an enhancement in the very low momentum region.
This result suggests that a systematic study of the multi-charm production can provide information about the charm distribution in momentum region hardly accessible trough observables derived from other charmed hadrons, i.e. \Dzer and \Lc; providing in this way a further observable able to infer charm quark interaction in the hot QCD matter.

\begin{center}
\begin{table} 
\begin{tabular}{c|c|c|c|c} 
\hline
\hline 
 &\Dzer & $\Lambda_{c}$ & $\Xi_{cc}^{+,++}$ & $\Omega_{ccc}$  \\ 
\hline 
\hline
$OO$         & 0.156   & 0.0732 & $3\!-\!12.1\!\cdot\!10^{-5}$  & $2.2\!-\!29.2 \!\cdot \!10^{-8}$ \\ 
$ArAr$       &0.543    & 0.301 & $1.9\!-\!6.6\!\cdot\!10^{-4}$  & $2.5\!-\!26.3\! \cdot\!10^{-7}$  \\ 
$KrKr$       &1.564    & 0.835 & $0.78\!-\!2.6\!\cdot\!10^{-3}$ & $1.5\!-\!14.9\! \cdot\!10^{-6}$ \\ 
$PbPb$       &5.343    & 3.0123   & $4\!-\!12.5\!\cdot\!10^{-3}$& $0.12\!-\!1.01\! \cdot\!10^{-4}$\\ 
\hline 
\end{tabular} 
\caption{\Dzer, \Lc, \Xicc and \Occc hadron $dN/dy$ in $OO$, $ArAr$, $KrKr$ and $PbPb$. Upper values corresponds to thermal charm \pt distribution, lower cases to realistic ones. See text for details.   }\label{Table:Yields}
\end{table}  
\end{center}
\section{Conclusions}\label{section:Conclusion}
In this paper, we have studied the single-charmed and multi-charmed hadron production by using an hadronization model by coalescence manly focusing on \Xicc and \Occc that are new heavy baryons likely to be detected in the next ALICE3 experiment at LHC. In particular, we have discussed the HF production in different collision systems, $PbPb$, $KrKr$, $ArAr$ and $OO$. We have considered in our study both realistic ("dynamical") charm distribution function coming from the evolution in QGP described by Boltzmann Transport Equation and thermal distribution with flow coming from the assumption of a thermalized source for charm quarks in order to have a better comparison to SHM. 
For \Dzer, $D^+$, $D^*$ and \Lc we have found a scaling with $A$ of the colliding system that is quite similar to SHM, but with absolute yield of \Lc, \Xic and $\Omega_c$ that are quite larger for the coalescence model. A result for \Pb that is in line with what found in $pp$ collisions \cite{Minissale:2020bif}. We have found that the charmed hadron yields in \Pb show a mass ordering with an enhancement for single-charmed baryons. We have found
that for D mesons and \Lc our model has compatible results with SHM in the case where the latter includes an enhanced set for baryons resonances. We predict yields for multi-charmed hadrons that is of the order of $10^{-2}\!-\!10^{-3}$ for doubled charmed hadrons, while is nearly two order of magnitude smaller in the case of \Occc. When we consider thermal distribution form charm quark we obtain a production that is comparable with the one from SHM.  
Moreover, we have studied the production of all hadrons changing the system size. 
Going from large to smaller collision systems we obtain roughly a decrease of the production of single-charmed and multi-charmed hadrons scaling with volume and number of charm as $N_c(N_c/V)^{C-1}$. But, focusing on the \Occc, this study shows a breaking of this simple scaling due to the change in shape of the charm quark distribution when a realistic simulation of the QGP medium is performed; suggesting that, looking at the yields, the \Occc production is an observable sensitive to the non-equilibrium features of charm quarks.  
In particular, we have seen that the ratio of \Occc \pt-spectra between different system size can provide a solid signature of a 
lack of thermalization when going to smaller system like O-O.
Furthermore, in PbPb we have seen a strong sensitivity of \Occc yield to its mean sqaure radius, at variance with D mesons and \Lc whose yields are mainly constrained by charm conservation.

\subsection*{Acknowledgments}
S.P. acknowledge the funding from UniCT under 'Linea di intervento 3' (HQsmall Grant).Y.S. thanks the sponsorship from Yangyang Development Fund. V.G. acknowledge the funding
from UniCT under 'Linea di intervento 2' (HQCDyn Grant).
This work was supported by the European Union’s Horizon 2020 research and innovation program Strong 2020 under grant agreement No. 824093.
We are thankful to F. Antinori,  A. Dainese and M. van Leeuwen for stimulating and fruitful discussions and comments.

\end{document}